\documentclass[9pt]{article}
\usepackage[T1]{fontenc}
\usepackage{graphicx}
\usepackage{bookman}
\usepackage{latexsym}
\usepackage{amssymb}
\usepackage{amsbsy}
\usepackage{cite}
\begin{document} 
  \title{Tensorial perturbations in an accelerating universe}
 \author{ M.de Campos$^{(1)}$}
\maketitle

%
%
   \begin{abstract}
 We study tensorial perturbations (gravitational waves) in a universe with particle production (OSC).  

The background of gravitational waves produces a perturbation in the redshift observed from distant sources.  
The modes for the perturbation in the redshift( induced redshift ) are calculated in a universe with particle production.
\\
\\
Key words - gravitational waves, perturbations, induced redshift.
\end{abstract}
\section{Introduction}
The inflationary paradigm indicates that we live in a flat universe, 
but observations from the cosmic microwave background anisotropy detected
only $30 \% $ of the material content necessary for the density parameter to reach the unity.
The non-luminous matter around  the galaxies is insufficient to complete
the deficit of $70 \% $ of the material content of the universe. 

The recent measurements on magnitude and redshift of  supernovae type Ia, 
made independently  by Perlmutter et   al. (1999)\cite{Perlmutter} and
Riess et al. (1998)\cite{Riess}, indicates that our universe is accelerating.
  Generally, this  acceleration is thought to be due
to any kind  of  repulsive gravity  , which can   be introduced via  a
negative pressure in the perfect fluid representation of the  universe.  So, some
candidates   to  this kind of  energy  are   discarded, namely: neutrinos,
kinetic energy, radiation.  All described by a positive pressure.

To supply the  energetic deficit of the universe perhaps
we live in a universe dominate by
a cosmological constant    \cite{Carrol} or a scalar field component, denominate  quintessence
\cite{Caldwell},\cite{Caldwell1}, \cite{Wang}. The quintessence models are an 
advantageous alternative to the cosmological constant, due the relation
with the super symmetric  models  \cite{Binetruy}, the problem  of the
fine-tuning of the cosmological constant \cite{Kolda} and supergravity
models \cite{Brax}. 

On another hand, a negative pressure can be provided in the perfect fluid energy momentum tensor, 
taking into account a cosmological particle production.  The   cosmological scenario  with particle
production, denominated by open system cosmology (OSC), was introduced by   Prigogine et. al \cite{Prigogine}  with
the intention of solving the problem of entropy content in the universe.
Note that, Einstein's equations can hardly provide  an explanation for the origin
of the cosmological entropy, since they are purely adiabatic and reversible. 

Traditionally the variables describing  the cosmological fluid are the
energy-density ($\rho$) and the  themodynamical pressure ($P_{th}$).
In  OSC a supplementary variable, the  particle number density, enters
into the description via a continuity equation with a source of
particles.  Naturally, the pressure in the stress energy
tensor is reinterpreted and a pressure due to the particle creation is
added to the thermodynamical pressure, called pressure creation ($P_c$).

Considering the source of particles  proportional to the product
of the number density and Hubble function, the prediction of the actual value
for the rate of the particle production  is  $10^{-16}nucleons/cm^3/Yr$ \cite{Lima1}. 
It is nearly the rate predicted by steady-state model  \cite{Hoyle}. 
Otherwise, it is far below the  detectable limit, the consequences for the dynamics 
of the universe can be evaluated.
OSC  can produce  solutions  for the very early  universe,
free of the   singularity  \cite{Abramo},  \cite{Prigogine}.   Besides
that, it can   generate  inflationary  models  without   the difficulties
associated   with   supercooling \cite{Lima}   and lead to a universe
sufficiently old as to agree with observations \cite{Alcaniz} \cite{Campos}.  

So, we consider the OSC
framework an promising background, from the observational
point of view, to study cosmological perturbations.

The interest in   the study of  cosmological tensorial perturbations
has increased  due to the  realization  that  they contribute   to   the
anisotropy of  the cosmic microwave background radiation \cite{Atrio},
\cite{Turner}, \cite{Turner1}, \cite{Grishchuk},\cite{Grishchuk1}. 

Gravitational waves provide   a  new  perspective of   the  observable
 universe.   The energy  density  of  gravitational waves can   affect
 matter,  however   weakly,  causing density   fluctuations,  peculiar
 velocities and tidal distortions, for   example.  On the another hand,  a
 cosmological  background   of   gravitational waves   induces redshift
 perturbations and angular deviations  in light transversing it.  The
 observations, at present, do not distinguish among anisotropies due
 density  perturbations and those produced   by gravitational waves at
 recombination era.  

In this work we solve the differential equation for tensorial modes
in a synchronous gauge for a cosmological background with particle 
production.

The redshift fluctuations ($\tilde{Z} $) induced by the gravitational waves
in OSC background  for an accelerated and non-accelerated universe 
are  calculated.
\section{Background}
\subsection{Open system cosmology}
In this section I attempt to give a summary of OSC basic equations.  The universe will
be  considered homogeneous and isotropic,   described by the FRW  line
element 
\begin{equation}
ds^2 = R^2(\eta)\{d\eta ^2-(dr^2+r^2 d\theta ^2+r^2 \sin{\theta}^2d\phi ^2)\}\, .
\end{equation}
The Einstein field equations for the  space-time described by the line element above and
the usual energy momentum tensor of an ideal fluid with an additional pressure \cite{Prigogine} due to 
the particle production ($P_c $)  are given by
\begin{eqnarray}
3(\frac{{R}^{\prime}}{R})^2 = \kappa R^2 \rho \\
-2\frac{{R^ {\prime \prime }}}{R} - \frac{{R}^{\prime}}{R} = \kappa R^2 P = \kappa R^2 (P_{th}+P_c) \, ,
\end{eqnarray}
where the prime means derivative with respect to the conform time $\eta $, 
and  $\kappa  = 8\pi  G$, $\rho $ is the energy density, $P_{th}$ is the thermodynamical pressure, 
and $P_c $ is the creation pressure.

The field equations (2) and (3) are coupled with the balance equation for the 
particle number density
\begin{equation}
{n}^{\prime}+n\theta = \Psi \, ,
\end{equation}
where $\theta = 3\frac{{R ^ {\prime}}}{R}$ and $\Psi$ is the source of
the particle production. 

The process of transference of the energy from the gravitational field to the production of particles acts as 
a source of entropy.  Using Gibbs relation, the energy conservation law ($u^{\alpha} {T^{\alpha \beta }} _{; \beta } = 0$, where $u^{\alpha }$
is the velocity four-vector)
 and balance equation for the particle number density, it is readily obtained that
\begin{eqnarray}
{S^{\alpha }} _{; \alpha } = n \dot {\sigma } + \sigma \Psi = \frac{ - P_c \theta}{T}-\frac{\mu \Psi }{T} \, , \nonumber
\end{eqnarray}
where $\sigma $ is the entropy per particle, $S^{\alpha }$ is the entropy current,  $\mu = \frac{\rho + P_{th}}{n} - T \sigma$
is the chemical potential, $T$ is the temperature, and the dot means the usual time derivative.  Considering the process as adiabatic ( $\dot{\sigma} = 0$ ) the pressure creation is given by
 \cite{Calvao}
\begin{equation}
P_c = -\frac{\rho +P_{th}}{n\theta}\Psi \, .
\end{equation}
Combining equations (2), (3), (4) and the state equation
\begin{eqnarray}
 P_{th} = \nu \rho\, , \nonumber
\end{eqnarray}
where $\nu $ is a constant, it follows that 
\begin{equation}
R{ R^{\prime \prime}} + [(\frac{3\nu +1}{2})- \frac{(\nu +1)\Psi}{2nH}] {R}^{\prime 2} = 0 \, .
\end{equation}
In order to  integrate equation (6) we write out an expression for the source $\Psi $.
Following Lima et al. \cite{Lima}, a physically reasonable expression for the particle creation rate
is $\Psi = 3 n \beta H$. To have the characteristic time for matter creation the Hubble time itself, 
 $\beta $ must be constant.  Beyond this, $\beta $ must be a positive constant, otherwise the second law of thermodynamics (${S^{\alpha}}
 _ {; \alpha}> 0 $)
is violated.  So, integrating (6) the expansion of the universe is governed by
the scale factor
\begin{equation}
R=\{R_0 (\frac{\eta}{\eta_0})\}^{\frac{2}{3(1-\beta)(\nu+1)-2}} \, ,
\end{equation}
where
\begin{equation}
R_0 = {{{\frac{3(1-\beta)(\nu+1)-2}{2}}(\frac{N_0 ^{\nu +1}m}{3})^{\frac{1}{2}}\eta _0}}^{\frac{2}{3(1-\beta)(\nu +1)-2}} \, .
\end{equation}
The subscript  $0 $ is  relative to the present time, $m$ is the  mass of
produced particle and $N $ is the particle number. 

The universe that evolves according the scale factor (7) emerges from a singularity.  
However, in the work of the Prigogine et al. \cite{Prigogine}, the source 
of the particles is proportional to the square of the Hubble function
 and the universe emerges from an initial instability given by an initial number 
of particles, instead of a singularity.
\subsection{Acceleration of the universe}
According to data from type Ia supernovae observations, the universe is
accelerating\cite{Goobar}. Generally, a negative pressure is
responsible for the increase of the expansion velocity of the universe and can be generated
from the inclusion of the cosmological term \cite{Carrol} or using a quintessence
component \cite{Signore}, for example. 

An alternative view would be to consider particle creation, which naturally
redefine the energy-momentum tensor and could account for this increasing
expansion velocity.   
Taking into account (6), we can write the deacceleration parameter\cite{Weinberg}  in terms
of the particle source function, namely
\begin{equation}
q=\frac{1}{2}-\frac{4\pi m G \Psi}{H^3}\, ,
\end{equation}
where $m $ is the rest mass of the  produced particles, and we consider a
null value for the curvature.  Considering $\Psi = 3n \beta H $, we obtain an
accelerated universe, expanding with the scale factor (7), if $\beta >
\frac{1}{3}$ \cite{Lima1}.
\section{Tensorial modes}
The background  is described by  the  line  element  (7).  In order  to
obtain the perturbed field  equations for tensorial  modes, we must to
return to    the  original field    equations,    substituting
$\tilde{g}_{\mu \nu} = g_{\mu \nu} + h_{\mu  \nu}$, where $g_{\mu \nu}
$  is the background  solution  and  $h_{\mu  \nu}$ is a  perturbation
around $g_{\mu   \nu}$.     The  perturbed field equations  are  
\cite{Weinberg} 
\begin{equation}
\delta R_{\mu \nu} = -8\pi G \delta S_{\mu \nu} \, ,
\end{equation}
where $S_{\mu \nu} = T_{\mu \nu } - \frac{1}{2}g_{\mu \nu }{T^{\lambda}} _{\lambda}$, and $T_{\mu \nu}$
is the energy momentum tensor.
 Therefore, the equation for evolution of the tensorial modes
in a synchronous gauge is given by \cite{Fabris}
\begin{equation}
 h^{\prime \prime}-2\frac{R^{\prime}}{R}h^{\prime}+[n^2 -2(\frac{R^{\prime \prime}}
{R}-\frac{R^{\prime 2}}{R ^2})]h=0 \, ,
\end{equation}
where the  spatial dependence  of the  perturbed  quantities are given  in
terms of spherical harmonics
 \begin{eqnarray}
\nabla ^2 h_{ij}(x,  \eta) = -h(\eta) n^2 Q_{ij}\, . \nonumber
\end{eqnarray}
$Q_{ij} $ is tracelles transverse eigenfunction in 3-d spatial section, so that $\nabla ^2 Q_{ij}=-n^2 Q_{ij}$.

Substituting the scale factor   (7) in equation   (11), we find  
\begin{equation}
h_a = C_1\eta^{k_1}J_w (n\eta)\, ,
\end{equation}
\begin{equation}
h_b = C_2\eta^{k_1}Y_w (n\eta) \, ,
\end{equation}
where $C_1$ and $C_2$ are integrations constants,
\begin{equation}
k_1 = \frac{1}{2} \frac{3(\beta -1) (\nu +1) -2}{3(\beta -1) (\nu +1) +2}
\end{equation}
and
\begin{equation}
w = \frac{3}{2}\frac{\beta (\nu +1)- \nu +1}{3\beta (\nu +1)-3\nu -1} \, .
\end{equation}
The functions $J$ and $Y$ are the Bessel functions of the first and second kind respectively.

The evolution  of the tensorial modes (12) and (13) will depend of the
parameter $\beta $.   Consequently, will depend on the deacceleration parameter (9). 
\newpage
For an accelerated matter dominated universe ($\beta > \frac{1}{3}, \nu =0$), we have oscillatory decaying modes (Fig. 1 ). 
\begin{figure}[!ht]
\centerline{\includegraphics[width=6cm]{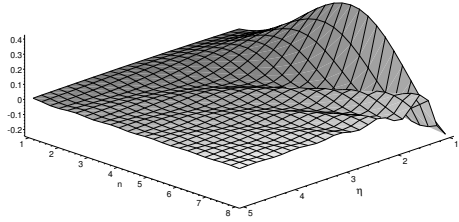}}
\caption{Evolution of the mode (12) for $\beta = \frac{5}{9}$ and $\nu =0 $.}
\label{fig:figure 1}
\end{figure}

On the other hand, for a non-accelerated universe we have oscillatory growing modes (Fig. 2 ).
\begin{figure}[!ht]
\centerline{\includegraphics[width=6cm]{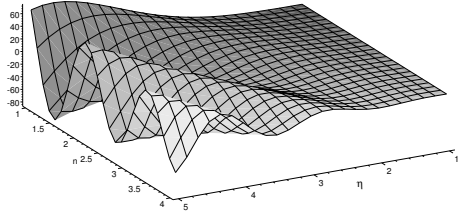}}
\caption{Evolution of the mode (13) for $\beta = \frac{1}{9}$ and $\nu =0 $.}
\label{fig:figure 2}
\end{figure}
For the  radiation dominated era the profiles   of the tensorial modes
are similar to the matter dominated era.

The integration, for the large scale perturbations ($n \rightarrow 0$),  of the equation (11),
furnishes the growing mode: 
\begin{equation}
h_1(\eta ) = C_1 \eta
\end{equation}
and a mode dependent of the creation parameter
\begin{equation}
h_2(\eta )=  C_2 \eta ^ {\frac{-4}{3 \beta (\nu +1)-3 \nu -1}} \, .
\end{equation}
From observational point of view the large scale perturbation has some advantages.  The measure of
the anisotropy of cosmic microwave background radiation is well established for $n \rightarrow 0$, that is, for small value of $l$
, where $l $ denotes the multipolar order in the expansion of the two point correlation function of temperature
\begin{eqnarray}
C(\theta) = \Sigma ^{\infty} _{l=2} C_l P_l (\cos (\theta)) \, . \nonumber
\end{eqnarray}

Using $0< \beta < \frac{1}{3}$ the universe is non-accelerated and the mode (17) will behave as 
a growing mode.
\section{Induced redshift}
How the propagation of light is affected by the less than perfectly symmetric and smooth universe
(here gravitational waves) is our goal in this section.
Basically, gravitational waves effects can be put in two categories.
Direct ones involve energy density of the gravitational waves affecting matter (for example, density fluctuations, peculiar velocities, 
tidal distortions) and induced effects that act on the propagation of radiation.
We calculate the induction,  by a
cosmological  background of  gravitational waves, of  redshift
perturbations  in light transversing  it in a universe with particle production.  The induced effect  on the
reshift is given by the relation \cite{Linder} \cite{Linder1}
\begin{equation}
Z = \frac{R_0}{R}-1+\tilde{Z} \, ,
\end{equation}
where
\begin{equation}
\tilde{Z}=-\frac{R_0}{2R}{\int}^{\eta }_{0} \frac{\partial h_{11}}{\partial \eta} d \eta \, .
\end{equation}

Writing   $h_{ij}=F(t)\exp{i\vec{k}\centerdot        \vec{x}}$, where
$\vec{x}$ is the comoving distance and $ \vec{k}$ is the comoving wave
vector,  given   by $ k=   \frac{2\pi  R}{\lambda}$, and considering the
large wavelength   perturbations, we obtain the induced redshift for the 
tensorial  perturbations (16) and (17), respectively:
\begin{equation}
\tilde{Z}_{1}=C_1 \eta ^{\frac{3(1-\beta )(1+\nu )-4}{3(1-\beta )(\nu +1)-2}}
\end{equation}
and
\begin{equation}
\tilde{Z}_{2}=C_2 \eta ^{\frac{2}{3(1-\beta)(\nu +1)-2}}\, .
\end{equation}
The induced redshift is a first order effect in the amplitude perturbations,  altering
our perception of the universe, not the physical constituents themselves.
Note that these perturbations in the redshift are dependents of the creation parameter $\beta $, 
consequently these modes are affected by the acceleration of the universe.

Similarities of this effect with gravitational lensing was point out by E. Linder \cite{Linder1}
\section{Conclusions}
The particle production in the universe results in a reinterpretation of the energy momentum tensor.
  The ordinary pressure is modified and divided in  thermodynamic
pressure and a negative pressure due to the creation of particles.  

The cosmological particle creation furnishes an alternative model
to explain the increasing of the expansion velocity of the universe and furnishes an suitable estimative for the age
of the universe. Taking into account the scale factor (7) the Hubble function is given by
\begin{eqnarray}
H = \frac{2}{3(1-\beta)(1+\nu)t}\, . \nonumber
\end{eqnarray}
If we do not violate the dominant energy condition and the second law of thermodynamics the constant $\beta$
stay in the interval $0 \leqslant \beta < 1$. Note that, for this interval of $\beta $, the universe will be
older than the universe described by standard model without creation ($\beta = 0$). Consequently, the conflict between
the age of the universe and the age of the oldest stars in our galaxy is eliminate in OSC \cite{Campos}, \cite{Chaboyer}.
So, we consider OSC an promising background from the observational point of view to realize perturbations.

We obtain the tensorial modes, for large wavelength perturbations and modes
dependent on the scale of the perturbations, considering the OSC as a background.
The behavior of these modes are sensitive to the creation parameter and consequently to the dynamics of the universe.
Taking into account $\beta = 0$ we obtain the usual solution for the ordinary fluid
without particle production.

The induced redshift ( $\tilde{Z} $) due to the cosmological gravitational waves are determined for large wavelength
perturbations.  The induced effects are small, but potentially observable.
The induced redshift can be , in principle, used to clarify the dynamics of the expansion of the universe, include for redshift 
 $z >> 1$.

Analyzing the induced redshift expressions,eq. (20) and eq. (21), we conclude that,  if we live in a deaccelerated universe
 ($\beta < \frac{1}{3}$) eq. (20) is a decaying mode and eq. (21) a growing mode.
Otherwise, for an accelerated universe  ($\beta >\frac{1}{3} $) we obtain the inverse behaviour for the  induced redshift.
So, the induced redshift for more distant objects is greater than the close objects, this seems  plausible. 

The growing mode for the induced redshift (eq.(21)) can contributes for give us an impression that the observed object
has a redshifit greater than the expected in a universe without the background of tensorial perturbations,
 even than we lived in a deaccelerated universe ($\beta < \frac{1}{3}$). 

The contribution of gravitational waves, generated in a universe with particle creation, in the anisotropy of the cosmic microwave 
background radiation is subject for a future study.

\section{Acknowledgments}
I like to acknowledgment to the Brazilian agency (CNPQ) for the financial support.

\end{document}